\documentclass[sigconf,nonacm]{acmart}

\usepackage{booktabs}
\usepackage{multirow}
\usepackage{graphicx}

\AtBeginDocument{}
\setcopyright{none}

\begin{document}

\title{GPU-Resident Speculative Decoding for In-Browser LLM Inference with WebGPU}

\author{Wei Song}
\affiliation{%
  \institution{Griffith University}
  \country{Australia}}

\author{Yuxin Cao}
\affiliation{%
  \institution{National University of Singapore}
  \country{Singapore}}

\author{Ziqi Ding}
\affiliation{%
  \institution{University of New South Wales}
  \country{Australia}}

\author{Xiao Cheng}
\affiliation{%
  \institution{Macquarie University}
  \country{Australia}}

\renewcommand{\shortauthors}{Song et al.}

\begin{abstract}
WebGPU enables private, offline LLM inference inside web pages, but browser dispatch, inter-process communication, synchronous readback and single-token kernels slow decoding and make conventional speculative decoding inefficient. We present a WebGPU runtime that addresses these browser-specific costs. Deterministic small-batch integer kernels accelerate multi-token verification while preserving exact agreement with greedy decoding in the same quantized runtime. A GPU-resident loop performs drafting, verification, acceptance and KV-cache maintenance without a per-token CPU decision. A draft-free context-mixing drafter reuses page content, conversation history and session memory, while a device-aware controller adapts the speculation width and the submission window to measured costs. Across seven models from 0.5B to 8B parameters, 11 tasks and three client GPUs, our runtime decodes 3.5 to 5.4$\times$ faster than WebLLM on a laptop GPU, 4.8 to 6.2$\times$ faster on an integrated GPU and 4.8 to 8.6$\times$ faster on a workstation GPU. The GPU-resident loop and speculation contribute a further 1.8 to 3.2$\times$ over the optimized kernels on the workstation. Outputs match the quantized greedy decoding of the runtime exactly, and the mean GSM8K accuracy matches that of WebLLM. Optional draft heads raise the peak speedup to 10.2$\times$, while session memory provides additional gains without downloading extra weights.
\end{abstract}



\maketitle

\section{Introduction}

Language models are moving into the web page itself. With WebGPU~\cite{webgpu-spec}, a page can download a model of a few gigabytes once and run it on the GPU of its visitor, and engines such as WebLLM~\cite{webllm} already serve models up to 8B parameters in a browser tab. The page, the forms and the conversation then stay on the device, the site pays nothing per token, and assistants that summarize the open page or help inside web editors keep working offline.

Decoding is the bottleneck of these in-page assistants. A browser places a security boundary between the page and the GPU, so every WebGPU dispatch is validated in the renderer and forwarded to a separate GPU process, and every token that JavaScript reads back stalls the pipeline until the GPU drains. The kernels of current engines add a third cost, since they are compiled for one token per step and process several tokens through their prefill path. On a laptop GPU, WebLLM spends 20.15~ms per token on Qwen2.5-0.5B, while our kernels need 5.93~ms with the same per-token readback and 4.22~ms without it. Speculative decoding~\cite{leviathan2023fast,chen2023accelerating}, the standard remedy for memory-bound decoding on servers, works against the platform in its usual form. A draft model pays the dispatch overhead of its own forward passes, the prefill kernels of WebLLM make the verification of 8 tokens cost 2.7 times a decoding step, and a CPU that inspects every verification result pays the full cost of the readback again at every step. The effect is strong enough to reverse the usual recipe. A fixed tree of 8 nodes, which speeds up decoding by half on a workstation GPU, makes it 2.7 times slower than plain decoding on the integrated GPU of a laptop.

This paper argues that the whole decoding loop of a browser runtime, from the kernels to the control of speculation, must be designed around the costs of the browser, and it presents a runtime that does so. To our knowledge, it is the first WebGPU runtime that moves tree-based speculative decoding, including drafting, verification, acceptance and KV-cache maintenance, into a GPU-resident loop that avoids per-token submission and readback in the browser. This requires deterministic small-batch verification kernels and control of the speculation width and of the submission window from costs that are measured on the device. Its small-batch kernels multiply 4-bit weights with 8-bit activations through integer dot products, so that verifying 8 tokens costs 19\% to 77\% more than one token on discrete GPUs instead of 2.7 times, and their reductions follow a fixed order, so that the verification of $N$ tokens is bit-identical to $N$ single-token steps and speculation reproduces the greedy decoding of the same quantized runtime exactly. The speculation loop runs entirely on the GPU. Drafting, verification, acceptance and the maintenance of the KV cache are kernels that read their control state from a GPU buffer, so the runtime encodes several steps into one command buffer and the CPU reads only a few bytes between windows of steps. A draft-free drafter in the spirit of context mixing in data compression~\cite{cleary1984ppm,mahoney2005paq} predicts continuations from the page, the conversation and a memory of earlier sessions, with no second model to download or run. A feedback controller measures the cost of each tree width on the device of the visitor and chooses the width that maximizes the expected tokens per millisecond, because a step of 8 nodes costs 19\% more than a step of one node on a workstation GPU and 4.1 times as much on the integrated GPU of a laptop.

We evaluate the runtime as a page would deploy it, in Chrome on the GPU of the visitor, with a laptop discrete GPU, an integrated GPU and a workstation GPU, using seven models from 0.5B to 8B parameters and 11 tasks, seven of which are grounded in web content. We compare against the three in-browser engines WebLLM, Transformers.js and LlamaWeb on all three GPUs. Controlled ablations separate the effects of the optimized kernels, GPU-resident execution and adaptive speculation, while cross-device tests characterize when wide verification helps or hurts. We also measure request latency, output agreement and task quality. Because the model shares the GPU with the page, we test whether the page still renders at 60 frames per second while it decodes, and we report the one-time cost that a page pays to load and calibrate a model, with the session memory kept in the same-origin storage of the page. These results establish both the performance benefits and the hardware-dependent limits of GPU-resident speculation in the browser. The contributions of this paper are as follows.
\begin{itemize}
\item We measure the costs of dispatch, readback and multi-token kernels that make in-browser decoding slow, and we show that the same costs make naive speculation slower than plain decoding on weak client GPUs.
\item We introduce GPU-resident speculative decoding for WebGPU, which removes the per-token acceptance decision of the CPU and its readback, and deterministic small-batch kernels that make multi-token steps cheap while keeping exact agreement with our quantized greedy decoder.
\item We propose a draft-free context-mixing drafter with a session memory, and a cost-aware controller that adapts the speculation width to each client GPU.
\item On three client GPUs, our runtime decodes 3.5 to 8.6$\times$ faster than WebLLM with output identical to its own greedy decoding and equal mean task accuracy, and ablations attribute the gain to each component of the runtime.
\end{itemize}

\section{Background and Motivation}
\label{sec:background}

\subsection{In-Browser LLM Inference}

WebGPU gives web pages direct access to the GPU through compute shaders written in WGSL~\cite{webgpu-spec}. Production runtimes such as WebLLM~\cite{webllm} compile a model into a few hundred WGSL kernels and drive them from JavaScript, and every token follows the same loop. The runtime encodes the kernels of one forward pass into a command buffer, submits it, waits until the logits or the sampled token reach JavaScript, and only then decides what to encode next.

\subsection{The Cost of a Token in WebGPU}

This loop is cheap on a native runtime and expensive in a browser, for three reasons that we measured on a laptop with an NVIDIA GTX 1650 in Chrome. Firstly, each dispatch passes through validation in the renderer process and an IPC hop to the GPU process, so a forward pass with 250 to 400 dispatches carries several milliseconds of fixed overhead~\cite{maczan2026dispatch}. Secondly, the per-token readback serializes the CPU and the GPU. In a microbenchmark with a fixed prompt, WebLLM spends 19.1~ms per token on Qwen2.5-0.5B with its synchronous loop, and the same kernels take 11.0~ms per token when we enqueue the steps back to back without waiting. This gap of 8~ms covers the whole host round trip of WebLLM per token, including its sampling and bookkeeping in JavaScript, whereas copying a sampled token back costs at most 1.9~ms in our runtime, as Section~\ref{sec:exp-e2e} shows. Thirdly, the kernels are tuned for one token per step. Processing several tokens per step goes through the prefill kernels of the runtime, and in WebLLM the verification of 4 and 8 tokens costs 1.75 and 2.7 times a single decoding step. All three costs work against speculative decoding, whose benefit is the number of accepted tokens per step divided by the cost of that step.

\section{Design}
\label{sec:design}

\begin{figure*}[t]
\centering
\includegraphics[width=\textwidth]{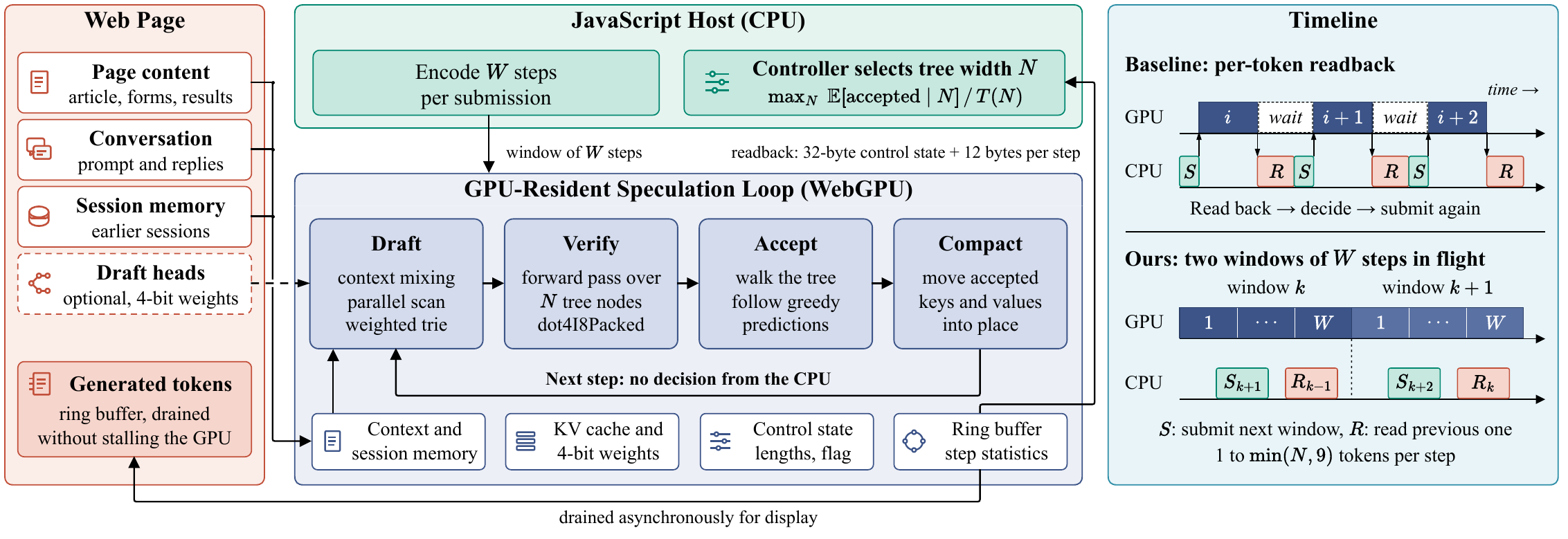}
\caption{Overview of the runtime. The GPU runs drafting, verification, acceptance and KV compaction as one loop, while the host keeps two windows of $W$ steps in flight and sets the tree width $N$ between windows.}
\label{fig:overview}
\end{figure*}

\subsection{Overview}

The system has four parts that remove the costs of Section~\ref{sec:background} one by one, and Figure~\ref{fig:overview} shows how they fit together. Small-batch kernels make the verification of several tokens cost only a fraction more than one decoding step. A GPU-resident speculation loop keeps drafting, verification, acceptance and KV-cache maintenance on the GPU, so the CPU only submits windows of steps and reads a few bytes back. A draft-free drafter predicts continuations from the page, the conversation and the earlier sessions of the user without downloading or running a second model. A feedback controller chooses the width of the speculation tree between windows from statistics that the GPU writes as a side effect of each step. The timeline in Figure~\ref{fig:overview} contrasts the per-token readback of a conventional runtime, which leaves the GPU waiting for every decision of the CPU, with two windows in flight, which keep the GPU busy.

\subsection{Small-Batch Verification Kernels}

Verification multiplies every weight matrix by a handful of token vectors, so it reads the same bytes as a decoding step and performs a few more arithmetic operations per byte. We keep the weights in 4-bit codes with one 16-bit scale per 32 inputs, the same granularity as the q4f16\_1 format of WebLLM, and quantize the activations of each token to 8-bit integers per block of 32 values before every matrix multiplication. A weight word then meets an activation word in two calls to the WGSL built-in \texttt{dot4I8Packed}, which compiles to the native 4-way integer dot product on recent GPUs, and the zero point is removed with a per-block correction term. Each thread keeps the accumulators of several rows and several tokens in registers, so one activation load serves several rows and one weight load serves every token in the step. The kernels are generated per weight shape and per bucket of 1, 2, 4, 8, 16 and 32 tokens, tuned once per GPU class, and the activation quantization is fused into the RMSNorm and SiLU kernels that precede each matrix multiplication. Attention splits the visible context into chunks that separate workgroups process and a second pass merges, and the query heads that share a KV head read each key and value once.

\subsection{GPU-Resident Speculation Loop}

Every step processes a fixed number $N$ of tree nodes, where node 0 is the token produced by the previous step and nodes 1 to $N-1$ form the draft tree. The state that a CPU would normally keep, namely the committed KV length, the number of valid nodes, the context length and a termination flag, lives in a small GPU buffer that every kernel reads at run time. A step consists of the forward pass over the tree, in which each node attends to the committed cache and to its ancestors through a bit mask, an acceptance kernel that walks the tree from the root and follows the child whose token equals the greedy prediction of its parent, a compaction kernel that moves the keys and values of the accepted path into place, and the drafter that writes the next tree. Because no step needs a decision from the CPU, the runtime encodes a window of $W$ steps into one command buffer and keeps two windows in flight. The CPU submits the next window before it reads the 32-byte control state and the 12 bytes of statistics per step of the previous one, which the same submission copies into a small mapped buffer, so the GPU never waits for a decision of the CPU. A step commits the accepted draft nodes and the token that the target model predicts after the last of them, so it emits between 1 and $N$ tokens, and at most 9 because the drafter extends a candidate by at most 8 tokens. Generated tokens accumulate in a ring buffer that the page drains asynchronously for display, without stalling the GPU.

\subsection{Draft-Free Drafting From Web Context}

In-browser assistants mostly answer about the page that the user is reading, and their outputs reuse phrases from that page, from the conversation and from earlier sessions. The drafter exploits this redundancy without a draft model. Following prediction by partial matching and context mixing in data compression~\cite{cleary1984ppm,mahoney2005paq}, it looks for earlier occurrences of the current suffix at every order up to 8 and blends the continuations of all orders into one distribution over the next tokens, weighting each order by its reliability so far. The simulation study of Appendix~\ref{app:accept} uses these online weights, and the GPU drafter approximates them with a fixed function of the match length. On the GPU, this becomes a parallel scan in which every context position computes the length of its match with the current suffix, a bitonic selection of the best positions per workgroup, and a single workgroup that merges the continuations of the selected positions into a weighted trie and emits the most probable prefix-closed set of nodes as the next tree. A memory region in the same buffer holds the outputs of earlier sessions, which a page can persist in IndexedDB, and the same pass scans it.

Where a site can afford a larger download, the runtime also accepts draft heads in the style of Medusa~\cite{medusa}. Four small heads read the final hidden state of the last verified token and predict the tokens two to five positions ahead over the 32,768 most frequent tokens of the vocabulary, which keeps their 4-bit weights between 99 and 151~MB for the 3B to 8B models. We train them by self-distillation on UltraChat~\cite{ding2023ultrachat} conversations and on 10.8K prompts built from the training splits of CNN/DailyMail, XSum, SQuAD, HotpotQA, MS MARCO and GSM8K, with the evaluation prompts held out, while HumanEval and MT-Bench are not part of their training data. A hybrid tree shares one verification step between the two drafters. The context drafter fills up to half of the nodes, or all of them when its best match is at least four tokens long, and the rest go to a best-first expansion of the head predictions ranked by the product of their probabilities. The heads add one small matrix product per step and require no change to the GPU-resident loop.

\subsection{Feedback Control of Speculation Width}

A wide tree pays off only when the drafts are good, and the quality of context drafts changes from one passage to the next. The runtime first measures the step cost $T(N)$ of every tree size at load time. During generation, the acceptance kernel records the number of accepted nodes of every step and the drafter records the match length of its best candidate. Between windows, the controller estimates the expected number of accepted tokens of each candidate width from the recent match lengths, using the acceptance that it observed for the same match length, and selects the width that maximizes the expected tokens per millisecond. Like a congestion controller, it occasionally probes a wider tree after a period of plain decoding, so that its statistics recover when the text becomes more predictable. The first width and the spacing of these probes also follow the measured costs, so a device on which wide steps are expensive opens with plain decoding and probes rarely. The costs are also refined online. Because two windows are in flight, the interval between the completions of two consecutive windows is the cost of the later window on the device as it runs at that moment, and the controller folds it into the cost of that width with an exponential moving average. This corrects a load-time measurement that interference or a change of clock has distorted, and it also captures the case in which the host limits the rate of steps. The window size follows the same principle. A larger window spreads the host work of a submission over more steps, but it commits a stale width for longer and runs past the end of a reply, so the runtime times a short decode with windows of one, two and four steps at load time and keeps the smallest window that is within 10\% of the fastest.

\section{Implementation}
\label{sec:implementation}

The runtime is a self-contained JavaScript module of about 1,500 lines that generate its WGSL kernels, and it runs unmodified in Chrome and Edge. We will release it as open source upon publication. It loads Qwen2.5 and Llama-3.2 checkpoints that an offline converter quantizes to 4-bit codes with one 16-bit scale per 32 inputs, and it keeps the KV cache in 32-bit floats. Kernel variants are generated from templates per weight shape and per token bucket, and a table of tuned configurations per GPU class selects among them, with a heuristic fallback for unseen shapes. A deterministic mode fixes the reduction order of every matrix multiplication and of attention across all buckets, so that the verification of $N$ tokens produces the same bits as $N$ single-token steps and speculative decoding reproduces the greedy decoding of the same runtime exactly. The engine uses this mode in every experiment unless stated otherwise. The deterministic mode costs little. On the workstation, the nondeterministic variant is not faster, and on the laptop GPU, it is 4\% to 15\% faster, but without the fixed order only 69 to 99 of the 220 speculative outputs per model equal the greedy output on the workstation and 81 to 94 on the laptop GPU, so exact speculation depends on the fixed order of reductions.

\section{Experiments}
\label{sec:experiments}

\subsection{Experimental Setup}

\noindent\textbf{Models.} We evaluate instruction-tuned Qwen2.5 (0.5B, 1.5B, 3B and 7B), Llama-3.2 (1B and 3B), Llama-3.1-8B, Gemma-2-2B and Phi-3.5-mini. The browser experiments use the first seven. The workstation runs all of them, and the two GPUs of the laptop serve as a lean cross-device check with the three smallest models, Qwen2.5-0.5B, Qwen2.5-1.5B and Llama-3.2-1B, which fit both. The acceptance study covers all nine models. All browser models use 4-bit weights with one scale per 32 inputs.

\begin{table}[t]
\centering
\caption{Engine configurations and completed prompts out of 220 for Qwen2.5-0.5B, Qwen2.5-1.5B and Llama-3.2-1B.}
\label{tab:baselines}
\small
\setlength{\tabcolsep}{2.5pt}
\begin{tabular}{llllr}
\toprule
Engine & Device & Weights & Activations & Completed \\
\midrule
\multirow{3}{*}{WebLLM} & Laptop & q4f16\_1 & 16-bit float & 220, 150, 202 \\
 & Integrated & q4f16\_1 & 16-bit float & 88, 51, 64 \\
 & Workstation & q4f32\_1 & 32-bit float & 220 each \\
\multirow{3}{*}{Transformers.js} & Laptop & 4-bit & 32 or 16-bit float & 219, 220, 220 \\
 & Integrated & 4-bit & 32-bit float & 220 each \\
 & Workstation & 4-bit & 32 or 16-bit float & 220 each \\
\multirow{3}{*}{LlamaWeb} & Laptop & Q4\_0 & 16-bit float & 220 each \\
 & Integrated & Q4\_0 & 16-bit float & 220 each \\
 & Workstation & Q4\_0 & 16-bit float & 220 each \\
Ours & All & 4-bit & 8-bit integer & 220 each \\
\bottomrule
\end{tabular}
\end{table}

\noindent\textbf{Workloads.} In-browser assistants mostly work on the page in front of the user, so 7 of the 11 tasks are grounded in web content. Four tasks apply different instructions to CNN/DailyMail news articles~\cite{cnndm}, namely summarization, rewriting as bullet points, extraction of entities as JSON and open question answering. The others are question answering over MS MARCO web search results~\cite{msmarco}, over SQuAD~2.0 passages~\cite{squad2} and over HotpotQA paragraphs~\cite{hotpotqa}, single-sentence summarization on XSum~\cite{xsum}, code completion on HumanEval~\cite{humaneval}, math word problems on GSM8K~\cite{gsm8k} and open-ended chat on MT-Bench~\cite{mtbench}. Page content is truncated to 1,536 tokens. The acceptance study uses 500 prompts per task (all 164 HumanEval and 80 MT-Bench prompts), and the browser experiments use 20 prompts per task, 220 per model. A session consists of the 220 prompts of one model in task order, and Section~\ref{sec:exp-heads} also evaluates a random order that mixes the tasks. An extended benchmark of 100 further prompts per task, or 60 for MT-Bench, 1,060 per model and disjoint from the first 220, repeats the workstation experiments at almost five times the scale in Appendix~\ref{app:big}.

\begin{table*}[t]
\centering
\caption{Decoding latency per output token (ms) on the prompts that WebLLM completes. A dash marks a model without a 4-bit Transformers.js export.}
\label{tab:e2e}
\small
\begin{tabular}{llrrrrrrrr}
\toprule
Device & Model & Prompts & Transformers.js & LlamaWeb & WebLLM & Greedy & GPU Loop & Ours & Speedup \\
\midrule
\multirow{3}{*}{Laptop} & Qwen2.5-0.5B & 220 & 40.56 & 20.84 & 20.15 & 5.93 & 4.22 & 3.75 & 5.4$\times$ \\
 & Qwen2.5-1.5B & 150 & 76.79 & 30.20 & 38.02 & 12.44 & 10.54 & 8.87 & 4.3$\times$ \\
 & Llama-3.2-1B & 202 & 31.09 & 22.86 & 24.51 & 8.98 & 7.44 & 7.05 & 3.5$\times$ \\
\midrule
\multirow{3}{*}{Integrated} & Qwen2.5-0.5B & 88 & 55.38 & 32.67 & 52.25 & 11.37 & 9.61 & 8.42 & 6.2$\times$ \\
 & Qwen2.5-1.5B & 51 & 99.32 & 66.09 & 143.15 & 29.43 & 27.97 & 29.70 & 4.8$\times$ \\
 & Llama-3.2-1B & 64 & 64.90 & 41.19 & 107.65 & 22.63 & 21.26 & 22.63 & 4.8$\times$ \\
\midrule
\multirow{7}{*}{Workstation} & Qwen2.5-0.5B & 220 & 27.03 & 16.04 & 12.17 & 4.59 & 2.13 & 1.44 & 8.5$\times$ \\
 & Qwen2.5-1.5B & 220 & 30.34 & 16.38 & 19.14 & 6.05 & 3.43 & 2.23 & 8.6$\times$ \\
 & Llama-3.2-1B & 220 & 9.00 & 11.49 & 8.85 & 4.15 & 1.86 & 1.60 & 5.5$\times$ \\
 & Qwen2.5-3B & 220 & -- & 18.46 & 25.42 & 7.80 & 4.76 & 3.50 & 7.3$\times$ \\
 & Llama-3.2-3B & 220 & 14.15 & 14.95 & 19.13 & 6.28 & 3.78 & 3.13 & 6.1$\times$ \\
 & Qwen2.5-7B & 220 & -- & 20.90 & 25.11 & 9.18 & 6.11 & 4.87 & 5.2$\times$ \\
 & Llama-3.1-8B & 220 & -- & 23.54 & 24.17 & 9.31 & 6.09 & 5.08 & 4.8$\times$ \\
\bottomrule
\end{tabular}
\end{table*}

\noindent\textbf{Hardware.} The browser experiments run in headless Chrome~155 on three GPUs that bracket the range of client devices. The laptop has an AMD Ryzen 5 5600H with an NVIDIA GTX 1650 (4 GB, Turing) and runs Windows 11, and its integrated AMD Radeon Vega GPU, which shares the system memory, serves as the second device. The workstation has an NVIDIA RTX PRO 5000 (48 GB, Blackwell) and runs Ubuntu 24.04, where Chrome reaches the GPU through Vulkan. We call the three devices the laptop, the integrated GPU and the workstation. Each experiment owns its device, and only one browser runs on a device at a time. Greedy outputs for the acceptance study are generated with PyTorch in bfloat16 from the original checkpoints on a separate server GPU.

\noindent\textbf{Baselines.} The main baseline is WebLLM~0.2.86~\cite{webllm}, the most widely used in-browser engine, with greedy decoding through its completion API, and Table~\ref{tab:baselines} lists the configuration of every engine. WebLLM shares our 4-bit codes with one scale per 32 inputs, with 16-bit scales on the laptop GPUs and 32-bit scales in its q4f32\_1 format on the workstation, which adds half a bit per weight, and it keeps activations in 16-bit floats on the two laptop GPUs and in 32-bit floats on the workstation, whose Vulkan driver does not expose 16-bit floats to WebGPU, while our kernels quantize activations to 8-bit integers. Multi-chunk prefill in WebLLM fails with a disposed-object error on long prompts, so we configure a single prefill chunk of 4,096 tokens and reset the chat state before every request. With this setting, WebLLM completes all 220 prompts of every model on the workstation, but it still fails on long prompts on the laptop GPUs. The failures concentrate on the longest page-grounded prompts, so the completed subsets are shorter than the full workload, and every comparison with WebLLM uses the identical subset of completed prompts for all engines. For the same reason, we leave the 3B models out of the laptop comparison, since WebLLM completes only 68 and 66 of their 220 prompts there. The second baseline is Transformers.js~3.8.1~\cite{transformersjs}, which runs ONNX models through the WebGPU backend of ONNX Runtime Web~\cite{onnxruntime}. Its 4-bit ONNX exports cover four of our seven models, all except Qwen2.5-3B, Qwen2.5-7B and Llama-3.1-8B. We use 16-bit activations for Llama-3.2-1B on the laptop and for Llama-3.2-3B on the workstation, whose 32-bit model fails to load, and 32-bit activations elsewhere, since the 16-bit models produce degenerate output on the other devices and models. We feed the prompt in chunks of 256 tokens because whole long prompts occasionally stall. The third baseline is LlamaWeb~\cite{levine2026llamaweb}, the WebGPU backend of llama.cpp, which we run in the browser through the prebuilt wllama~3.8.1 build that the LlamaWeb authors use, with Q4\_0 weights for all seven models on the workstation and for the three smallest models on the laptop GPUs. It needs 16-bit floats, which Chrome exposes on the workstation only with an additional Dawn flag that we also set for the 16-bit Transformers.js model there, and its tokenizer reproduces our token ids on all prompts. Inside our runtime, we compare four decoding modes on identical weights and kernels, namely plain greedy decoding with a CPU readback after every token, the GPU-resident loop without speculation, speculation with a fixed tree of 8 nodes and speculation with the adaptive controller. This separates the gain of the integer kernels, from WebLLM to Greedy, from the gain of GPU residency, from Greedy to GPU Loop, and from that of speculation, from GPU Loop to Ours.

\noindent\textbf{Metrics.} We report the decoding latency per output token, which divides the total decoding time of a model by the number of tokens it generates over all prompts, each prompt generating up to 128 tokens. For the page-grounded tasks, we also report the time to the first token and the total latency of a request. Other metrics are the mean number of tokens produced per verification step ($\tau$) and the number of prompts on which the speculative output equals the plain greedy output token for token. We call speculation lossless in this strict sense, as equality with the deterministic quantized greedy decoder of the same runtime. It is not equality with the original bfloat16 checkpoint or with WebLLM, whose outputs depart from ours after a few tokens because of quantization, so quality is measured separately as perplexity on wikitext-2~\cite{wikitext}, over identical token windows in both engines, and as accuracy on 200 GSM8K problems with answers of up to 384 tokens, where the last number of an answer must equal the reference. The cross-device comparison is repeated three times on a fixed subset of 66 prompts, six per task, with every engine warmed up before timing.

\subsection{End-to-End Decoding Latency}
\label{sec:exp-e2e}

Table~\ref{tab:e2e} reports the decoding latency of three external baselines, WebLLM, Transformers.js and LlamaWeb, and of three configurations of our runtime, using the prompts that WebLLM completes for each comparison. Both alternative engines run on all three GPUs, and Transformers.js lacks 4-bit exports for three of the workstation models. On the workstation, which runs all seven models, our runtime decodes 4.8 to 8.6$\times$ faster than WebLLM, and for the three smallest models, it is 3.5 to 5.4$\times$ faster on the laptop GPU and 4.8 to 6.2$\times$ faster on the integrated GPU. Every speculative output equals the plain greedy output of the same runtime on all 220 prompts of every model on all three GPUs. On the laptop GPU and the workstation, the alternative engines range from 2.2$\times$ slower to 1.4$\times$ faster than WebLLM. On the integrated GPU, they are faster than WebLLM, by up to 1.7$\times$ for Transformers.js and by 1.6 to 2.6$\times$ for LlamaWeb. Our runtime remains the fastest engine on every device and model, ahead of the fastest baseline by 1.8 to 8.5$\times$, and the margin is narrowest on the integrated GPU, at 1.8 to 3.9$\times$. The gain is largest for small models on the fast GPU, where the fixed costs of the browser dominate, and it shrinks to 4.8 to 5.2$\times$ for the bandwidth-bound 7B and 8B models. On the workstation, removing the per-token readback alone cuts the latency of Qwen2.5-0.5B from 4.59 to 2.13~ms, and speculation removes another third. On the laptop GPU, the same readback costs at most 1.9~ms per token against a computation of 4 to 11~ms, so the loop and speculation together add 1.3 to 1.6$\times$, and on the integrated GPU they add little, because wide steps cost too much there.

\begin{figure}[t]
\centering
\includegraphics[width=\columnwidth]{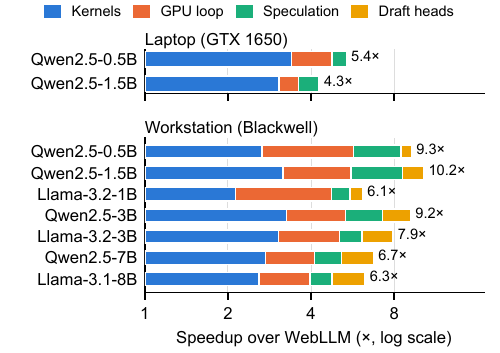}
\caption{Speedup over WebLLM decomposed by component.}
\label{fig:breakdown}
\end{figure}

Figure~\ref{fig:breakdown} attributes the speedup to the components of the runtime on the prompts that WebLLM completes, where the factors multiply to the total, for all seven models on the workstation and two representative models on the laptop GPU. The kernels contribute the most, at 2.1 to 3.3$\times$ on the workstation and 3.1 to 3.4$\times$ on the laptop GPU. This factor does not come from the 32-bit activations of WebLLM on the workstation, since it is at least as large on the laptop GPU, where WebLLM keeps its activations in 16-bit floats. The GPU-resident loop adds 1.50 to 2.23$\times$ on the workstation but 1.18 to 1.41$\times$ on the laptop GPU, and speculation adds 1.17 to 1.54$\times$ and 1.13 to 1.19$\times$, respectively. The optional draft heads of Section~\ref{sec:exp-heads} add up to 1.32$\times$ on top, so that Qwen2.5-1.5B decodes 10.2$\times$ faster than WebLLM on the workstation, while the default context-only runtime peaks at 8.6$\times$ in Table~\ref{tab:e2e}. The two components that are specific to the browser, the loop and the cost-aware speculation, add a further 1.8 to 3.2$\times$ on the workstation.

\begin{figure}[t]
\centering
\includegraphics[width=\columnwidth]{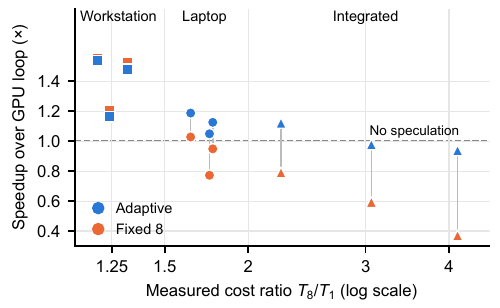}
\caption{Speculation speedup versus cost ratio.}
\label{fig:width}
\end{figure}

\subsection{Cross-Device Adaptive Speculation}
\label{sec:exp-width}

The value of speculation depends on how much a wider verification step costs, and this cost differs sharply across client GPUs. Figure~\ref{fig:width} plots the speedup of a fixed tree of 8 nodes and of the adaptive controller over the GPU-resident loop without speculation, against the cost ratio $T_8/T_1$, the median of five calibrations, for three models on each of the three GPUs. The numbers in this subsection use all 220 prompts for the comparison of policies within our runtime, while Table~\ref{tab:e2e} reports the subset that WebLLM completes. On the workstation, an 8-node step costs only 19\% to 32\% more than a 1-node step, so a fixed tree of 8 nodes is already near the best policy and the controller matches it within 4\%. On the laptop GPU, the same step costs 64\% to 77\% more, and on the integrated GPU it costs 124\% to 312\% more. The fixed tree then loses to plain decoding on five of the six configurations of these two devices, and it is 2.7$\times$ slower than the GPU loop for Llama-3.2-1B on the integrated GPU, at 61.02 against 22.59~ms, because the tokens that it gains do not repay the extra work of verification. The controller avoids most of this loss. It is the fastest of the three policies on the laptop GPU, with gains of up to 19\%, and for Qwen2.5-0.5B on the integrated GPU, with a gain of 12\%. For the two larger models on the integrated GPU, occasional exploration still leaves it 2\% and 7\% slower than the GPU loop without speculation, against 69\% and 170\% for the fixed tree, and about 6\% slower on the shorter prompts of Table~\ref{tab:e2e}, so the controller avoids severe slowdowns without guaranteeing a gain. A single runtime thus serves devices whose cost ratios differ by a factor of almost four without tuning per device, which matters on the web, where a page runs on whatever GPU the visitor owns. At 16 nodes, the step cost grows to 1.7$\times$ to 1.8$\times$ that of one node on the workstation and to 3.8$\times$ to 6.5$\times$ on the integrated GPU (Figure~\ref{fig:tn} in Appendix~\ref{app:tn}). Three runs on the 66-prompt subset differ by at most 4\% on the workstation, 2\% on the laptop GPU and 8\% on the integrated GPU, and the order of the three policies never changes. The chosen widths follow these curves. On the workstation, 41\% to 86\% of the steps use a tree of 8 nodes, on the laptop GPU no step does and 57\% to 66\% use 4 nodes, and on the integrated GPU, 84\% and 96\% of the steps of Qwen2.5-1.5B and Llama-3.2-1B use a single node, so the mean number of tokens per step falls from 1.45 to 1.80 on the workstation to 1.01 to 1.47 on the integrated GPU.

\subsection{Window Size and Responsiveness}
\label{sec:exp-window}

\begin{figure}[t]
\centering
\includegraphics[width=\columnwidth]{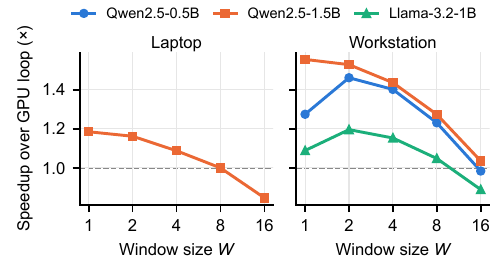}
\caption{Speedup against the window size.}
\label{fig:window}
\end{figure}

Figure~\ref{fig:window} varies the number of steps $W$ that the runtime encodes per submission, on all 220 prompts, with the adaptive controller and every output still exact. On the workstation, a window of two steps is fastest for Qwen2.5-0.5B and Llama-3.2-1B, at 1.45 and 1.58~ms, and a window of one step is fastest for Qwen2.5-1.5B, at 2.21~ms. Larger windows lose steadily, and at 16 steps the two smaller models decode more slowly than the GPU loop without speculation, because the controller commits a stale width for 16 steps and the window runs past the end of most replies. On the laptop GPU, a confirmation with Qwen2.5-1.5B finds a window of one step fastest, at 9.12~ms against 9.30, 9.93, 10.80 and 12.75~ms with two, four, eight and 16 steps. The calibration at load time searches only one, two and four steps, a subset of this sweep, and it picks exactly the fastest windows, two steps for the two smaller models on the workstation and one step elsewhere.

Small windows also keep the page responsive. A second WebGPU device in the same page submits a small compute pass every 16.7~ms, as an animation would, and records the time until each pass completes. While Qwen2.5-0.5B decodes on the workstation, the 99th percentile of this frame latency is 5.7~ms with a window of one step and 8.9~ms with two, against 2.9~ms on an idle GPU and 5.1~ms during plain greedy decoding, and the page keeps 60 frames per second. A window of 16 steps raises it to 42.7~ms and drops the page to 57 frames per second, because each frame waits behind a whole window. The laptop GPU keeps 60 frames per second in every configuration. The integrated GPU is different, since it shares its execution units and memory with the rest of the system. While Qwen2.5-0.5B decodes there, the page falls to 26 frames per second with a window of one step and to 35 frames per second even during plain greedy decoding, against 60 frames per second when idle, and a window of 16 steps drops it to 11 frames per second with a 99th percentile of 310~ms. Small windows are therefore necessary on this device but not sufficient, and a page that animates while it decodes on an integrated GPU has to throttle the decoder.

\subsection{Draft Heads and Session Memory}
\label{sec:exp-heads}

A replay simulation over the greedy outputs of all nine models in Appendix~\ref{app:accept} shows that our context-mixing drafter produces 1.83 tokens per step with 32 nodes, against 1.62 for prompt lookup and 1.73 for a suffix tree, and that session memory raises it to 2.40. Table~\ref{tab:heads} adds the two optional sources of drafts to the adaptive runtime on the workstation, where Context uses only the current request, Memory adds the outputs of earlier requests in the session and Heads adds the draft heads to the context drafter. Every configuration still produces the plain greedy output on all 220 prompts. The heads cut the latency by 9\% to 24\%, and for the 7B and 8B models, they reach 3.73 and 3.85~ms, which is 6.7 and 6.3$\times$ faster than WebLLM. Session memory reaches a large part of this gain without any download. It lowers the latency by 10\% to 17\%, and for Llama-3.2-1B, it is faster than the heads, at 1.40 against 1.44~ms. The memory holds the outputs that the runtime itself generated for the earlier evaluation requests of the same model, never reference answers, and it is cleared whenever a model is loaded. Benchmark requests may share more with each other than the requests of a real user, so we also run the 220 prompts in a random order that mixes all 11 tasks. Memory then changes the latency of Qwen2.5-0.5B, Qwen2.5-1.5B and Llama-3.2-1B by at most 0.02~ms, so the gain does not depend on sessions that repeat one task. The heads are trained on the training splits of six of our datasets with the evaluation prompts held out, so these tasks are in domain. On HumanEval and MT-Bench, which are out of domain, they cut the latency by 9\% to 25\%, as much as the 7\% to 24\% on the in-domain tasks.

\begin{table}[t]
\centering
\caption{Workstation latency (ms) with memory and heads.}
\label{tab:heads}
\small
\begin{tabular}{lrrr}
\toprule
Model & Context & Memory & Heads \\
\midrule
Qwen2.5-0.5B & 1.44 & 1.30 & 1.31 \\
Qwen2.5-1.5B & 2.23 & 1.95 & 1.87 \\
Llama-3.2-1B & 1.60 & 1.40 & 1.44 \\
Qwen2.5-3B & 3.50 & 3.03 & 2.77 \\
Llama-3.2-3B & 3.13 & 2.63 & 2.41 \\
Qwen2.5-7B & 4.87 & 4.30 & 3.73 \\
Llama-3.1-8B & 5.08 & 4.24 & 3.85 \\
\bottomrule
\end{tabular}
\end{table}

\subsection{Request Latency and Output Quality}
\label{sec:exp-quality}

Users experience whole requests rather than the latency of a token. Table~\ref{tab:request} breaks down the latency of the seven page-grounded tasks, whose prompts carry about 830 tokens of page content on the workstation, while the subsets that WebLLM completes on the laptop GPU are shorter. Our runtime reaches the first token 2.4 to 3.7$\times$ sooner than WebLLM, because its prefill runs the same integer kernels with the widest token bucket, while WebLLM runs floating-point prefill kernels over a single chunk, and it completes a request 4.8 to 7.1$\times$ faster on the workstation and 3.2 to 4.0$\times$ faster on the laptop GPU, where a Qwen2.5-0.5B answer about a news article takes 0.9~s against 3.2~s. These request speedups combine integer prefill with the decoding optimizations that Table~\ref{tab:e2e} isolates.

\begin{table}[t]
\centering
\caption{Request latency (ms) on the page-grounded tasks.}
\label{tab:request}
\small
\setlength{\tabcolsep}{3pt}
\begin{tabular}{llrrrr}
\toprule
\multirow{2}{*}{Device} & \multirow{2}{*}{Model} & \multicolumn{2}{c}{First token} & \multicolumn{2}{c}{Total} \\
\cmidrule(lr){3-4}\cmidrule(lr){5-6}
 & & WebLLM & Ours & WebLLM & Ours \\
\midrule
\multirow{3}{*}{Workstation} & Qwen2.5-0.5B & 366 & 124 & 1,494 & 249 \\
 & Qwen2.5-1.5B & 707 & 190 & 2,815 & 394 \\
 & Llama-3.2-1B & 401 & 125 & 1,462 & 302 \\
\midrule
\multirow{3}{*}{Laptop} & Qwen2.5-0.5B & 1,366 & 578 & 3,182 & 923 \\
 & Qwen2.5-1.5B & 2,937 & 841 & 6,302 & 1,569 \\
 & Llama-3.2-1B & 3,154 & 1,017 & 5,963 & 1,843 \\
\bottomrule
\end{tabular}
\end{table}

Faster kernels must not cost accuracy, and our kernels quantize activations to 8 bits where WebLLM keeps them in floating point. On wikitext-2, Qwen2.5-1.5B reaches a perplexity of 15.438 in our runtime and 15.426 in WebLLM, a difference of 0.08\%, and Llama-3.2-1B reaches 20.799 and 20.705, a difference of 0.45\%. Table~\ref{tab:gsm} measures task accuracy on GSM8K, where both engines run on the workstation with answers of up to 384 tokens. Over the seven models, our runtime and WebLLM solve the same share of problems, 52.3\% each. Individual models differ by up to 6\% in either direction, as for Qwen2.5-3B at 56.5\% against 62.5\% and Qwen2.5-0.5B at 23.0\% against 17.0\%, so the evidence supports parity on average rather than for every model. The paired difference over all 1,400 model-problem evaluations, 200 problems for each of seven models, is 0.0\% with a 95\% bootstrap interval from -2.2\% to 2.4\%, and the interval of every model includes zero. Both engines lose accuracy against bfloat16, mostly for the smallest models, through their shared 4-bit weights. The page-grounded tasks have no single reference answer, so we compare the output of each engine with the bfloat16 output for the same prompt by token-level ROUGE-L, for the six models whose bfloat16 outputs use the same prompts. Our runtime reaches 0.49 and WebLLM 0.46 on average. The two engines are within 0.02 of each other for five models, and for Qwen2.5-7B our outputs stay closer to the reference, at 0.57 against 0.43, so our 8-bit activations keep the outputs at least as close to full precision.

\begin{table}[t]
\centering
\caption{GSM8K accuracy (\%) on 200 problems.}
\label{tab:gsm}
\small
\begin{tabular}{lrrr}
\toprule
Model & PyTorch bfloat16 & WebLLM & Ours \\
\midrule
Qwen2.5-0.5B & 41.5 & 17.0 & 23.0 \\
Qwen2.5-1.5B & 61.0 & 50.0 & 54.0 \\
Llama-3.2-1B & 42.5 & 31.5 & 31.0 \\
Qwen2.5-3B & 66.5 & 62.5 & 56.5 \\
Llama-3.2-3B & 68.0 & 66.0 & 64.5 \\
Qwen2.5-7B & 68.0 & 70.0 & 69.0 \\
Llama-3.1-8B & 72.5 & 69.0 & 68.0 \\
\midrule
Mean & 60.0 & 52.3 & 52.3 \\
\bottomrule
\end{tabular}
\end{table}

\section{Discussion}
\label{sec:discussion}

\noindent\textbf{Privacy of Session Memory.} The session memory holds earlier outputs of the model, which can contain personal content. The runtime keeps it inside the origin of the page, in a GPU buffer during a visit and optionally in IndexedDB between visits, so the same-origin policy of the browser isolates it from other sites and clearing the site data removes it. The memory never leaves the device, and a page can disable persistence or scope it to one conversation.

\noindent\textbf{Deployment Cost.} A page pays a one-time cost when it loads a model. On the workstation, loading the 4-bit weights from a local server and compiling the kernels takes 3 to 11~s for the seven models, the calibration of the step costs takes 4 to 19~s, and the calibration of the window takes 2 to 6~s. Our speedups are steady-state and exclude these costs. A page could cache both calibrations in IndexedDB, which our prototype does not yet do, so against WebLLM a first visit recovers them only after 4 to 9 page-grounded requests for the three smallest models, and a single short request does not finish sooner. The KV cache of 2,560 tokens in 32-bit floats takes 63 to 671~MB, next to weights of 0.3 to 4.3~GB.

\noindent\textbf{Limitations.} The current design targets greedy decoding, since lossless sampling would require a verification rule that is deterministic under randomness. The experiments cover Chrome with the tested NVIDIA and AMD drivers, and the laptop results come from two GPUs of a single machine with the three smallest models. We have not measured Safari, Firefox or Apple and mobile GPUs, and we do not project our results to them. We compare against no native runtime and against no variant of our kernels with floating-point activations, so the kernel factor mixes the effect of 8-bit activations with that of the small-batch design.

\section{Related Work}
\label{sec:related}

\noindent\textbf{In-Browser and On-Device Inference.} Browser inference began with TensorFlow.js~\cite{smilkov2019tfjs}, several times slower than native execution~\cite{ma2019moving}. WebLLM~\cite{webllm} compiles language models with the machine learning compiler of TVM~\cite{chen2018tvm} into WGSL kernels, Transformers.js~\cite{transformersjs} runs models through ONNX Runtime Web~\cite{onnxruntime}, LlamaWeb~\cite{levine2026llamaweb} brings the portable kernels and static memory planning of llama.cpp to WebGPU, and bitgpu~\cite{bitgpu} verifies linear prompt-lookup drafts for 1-bit models. Recent work reduces the dispatch overhead of WebGPU by fusing kernels~\cite{maczan2026dispatch,maczan2026measuring}, WeInfer~\cite{chen2025weinfer} reuses buffers and pipelines the fetching of results on top of WebLLM, and kernels can be tuned just in time for the device~\cite{jia2024jit}. These systems accelerate one token per step, and WeInfer defers the readback of each token without removing it, whereas our runtime takes the per-token decision away from the CPU and makes speculation profitable. Native engines capture decoding and verification steps in CUDA graphs~\cite{kwon2023vllm}, and they never pay the validation and inter-process costs of a browser sandbox.

\noindent\textbf{Speculative Decoding.} Speculative decoding~\cite{leviathan2023fast,chen2023accelerating} verifies cheap drafts in one step of the target model. Tree-based verification checks several continuations at once~\cite{miao2024specinfer}, Medusa~\cite{medusa} and EAGLE~\cite{li2024eagle} replace the draft model with light heads on the target model, and Sequoia~\cite{chen2024sequoia} sizes the tree to the hardware. Dynamic speculation adapts the draft length from the confidence of the drafter~\cite{mamou2024dynamic,huang2025specdecpp} or the shape of the tree from draft probabilities~\cite{li-etal-2024-eagle}, and SmartSpec~\cite{liu2024smartspec} chooses the speculation length that maximizes the goodput of a serving system under a cost model. Our controller pursues the same objective between windows of GPU-resident steps, with costs measured on each client device. Lossless verification also needs kernels whose results do not depend on the number of tokens in a step, since a reduction order that depends on the batch size can change the token that greedy decoding selects.

\noindent\textbf{Draft-Free Drafting.} Prompt lookup decoding~\cite{saxena2023pld} copies the continuation of the last matching n-gram from the prompt, LLMA~\cite{yang2023llma} copies spans from reference documents, REST~\cite{he2024rest} retrieves drafts from a datastore, and lookahead decoding~\cite{fu2024lookahead} and Token Recycling~\cite{luo-etal-2025-turning} reuse candidates from earlier steps. SAM-Decoding~\cite{hu-etal-2025-sam} matches the suffix with a suffix automaton, and SuffixDecoding~\cite{oliaro2024suffix} drafts from suffix trees over earlier outputs. Our drafter blends the continuations of all match orders as in context mixing and runs as a parallel scan on the GPU, with no CPU index or readback. Our runtime implements the complete decision path of speculation as WebGPU kernels, unlike native graph execution or a CPU-driven speculation loop, and it enforces deterministic small-batch verification and device-aware submission under the validation, inter-process and synchronization constraints of the browser, giving exact speculation without a host decision per token.

\section{Conclusion}

Decoding in a browser becomes fast when the complete runtime is designed around the dispatch, synchronization and readback costs of the browser. Our runtime uses deterministic small-batch integer kernels, a GPU-resident decision path for speculation, draft-free context mixing from page and session content, and device-aware control of the speculation width. Across three client GPUs, it decodes up to 8.6$\times$ faster than WebLLM while it exactly reproduces the greedy decoding of the same quantized runtime. The cross-device results also show why a fixed speculation width is unsuitable for heterogeneous client hardware, since the controller limits severe regressions when wide verification is expensive. These findings establish GPU-resident, cost-aware execution as a practical direction for in-browser LLM inference, while first-visit startup and the responsiveness of integrated GPUs remain deployment constraints. Future work will study cheaper calibration on the first visit and more responsive scheduling on integrated and mobile GPUs, which share their execution units with the rest of the page.

\bibliographystyle{ACM-Reference-Format}
\bibliography{references}

@misc{webgpu-spec,
  title={{WebGPU}},
  author={Ninomiya, Kai and Jones, Brandon and Blandy, Jim},
  howpublished={W3C Candidate Recommendation Draft},
  url={https://www.w3.org/TR/webgpu/},
  year={2026}
}

@article{webllm,
  title={Webllm: A high-performance in-browser llm inference engine},
  author={Ruan, Charlie F and Qin, Yucheng and Parthasarathy, Akaash R and Zhou, Xun and Lai, Ruihang and Jin, Hongyi and Dong, Yixin and Hou, Bohan and Yu, Meng-Shiun and Zhai, Yiyan and others},
  journal={arXiv preprint arXiv:2412.15803},
  year={2024}
}

@inproceedings{leviathan2023fast,
  title={Fast inference from transformers via speculative decoding},
  author={Leviathan, Yaniv and Kalman, Matan and Matias, Yossi},
  booktitle={International Conference on Machine Learning},
  pages={19274--19286},
  year={2023},
  organization={PMLR}
}

@article{chen2023accelerating,
  title={Accelerating large language model decoding with speculative sampling},
  author={Chen, Charlie and Borgeaud, Sebastian and Irving, Geoffrey and Lespiau, Jean-Baptiste and Sifre, Laurent and Jumper, John},
  journal={arXiv preprint arXiv:2302.01318},
  year={2023}
}

@article{cleary1984ppm,
  title={Data compression using adaptive coding and partial string matching},
  author={Cleary, John and Witten, Ian},
  journal={IEEE Transactions on Communications},
  volume={32},
  number={4},
  pages={396--402},
  year={1984},
  publisher={IEEE}
}

@techreport{mahoney2005paq,
  title={Adaptive Weighing of Context Models for Lossless Data Compression},
  author={Mahoney, Matthew V.},
  institution={Florida Institute of Technology},
  number={CS-2005-16},
  year={2005}
}

@article{maczan2026dispatch,
  title={Characterizing WebGPU Dispatch Overhead for LLM Inference Across Four GPU Vendors, Three Backends, and Three Browsers},
  author={Maczan, J{\k{e}}drzej},
  journal={arXiv preprint arXiv:2604.02344},
  year={2026}
}

@inproceedings{medusa,
  title={Medusa: Simple LLM Inference Acceleration Framework with Multiple Decoding Heads},
  author={Cai, Tianle and Li, Yuhong and Geng, Zhengyang and Peng, Hongwu and Lee, Jason D and Chen, Deming and Dao, Tri},
  booktitle={International Conference on Machine Learning},
  year={2024}
}

@inproceedings{ding2023ultrachat,
  title={Enhancing chat language models by scaling high-quality instructional conversations},
  author={Ding, Ning and Chen, Yulin and Xu, Bokai and Qin, Yujia and Hu, Shengding and Liu, Zhiyuan and Sun, Maosong and Zhou, Bowen},
  booktitle={Proceedings of the Conference on Empirical Methods in Natural Language Processing},
  pages={3029--3051},
  year={2023}
}

@inproceedings{cnndm,
  title={Get to the point: Summarization with pointer-generator networks},
  author={See, Abigail and Liu, Peter J and Manning, Christopher D},
  booktitle={Proceedings of the Annual Meeting of the Association for Computational Linguistics},
  pages={1073--1083},
  year={2017}
}

@article{msmarco,
  title={Ms marco: A human generated machine reading comprehension dataset},
  author={Bajaj, Payal and Campos, Daniel and Craswell, Nick and Deng, Li and Gao, Jianfeng and Liu, Xiaodong and Majumder, Rangan and McNamara, Andrew and Mitra, Bhaskar and Nguyen, Tri and others},
  journal={arXiv preprint arXiv:1611.09268},
  year={2016}
}

@inproceedings{squad2,
  title={Know what you don’t know: Unanswerable questions for SQuAD},
  author={Rajpurkar, Pranav and Jia, Robin and Liang, Percy},
  booktitle={Proceedings of the Annual Meeting of the Association for Computational Linguistics},
  pages={784--789},
  year={2018}
}

@inproceedings{hotpotqa,
  title={HotpotQA: A dataset for diverse, explainable multi-hop question answering},
  author={Yang, Zhilin and Qi, Peng and Zhang, Saizheng and Bengio, Yoshua and Cohen, William and Salakhutdinov, Ruslan and Manning, Christopher D},
  booktitle={Proceedings of the Conference on Empirical Methods in Natural Language Processing},
  pages={2369--2380},
  year={2018}
}

@inproceedings{xsum,
  title={Don’t give me the details, just the summary! topic-aware convolutional neural networks for extreme summarization},
  author={Narayan, Shashi and Cohen, Shay B and Lapata, Mirella},
  booktitle={Proceedings of the Conference on Empirical Methods in Natural Language Processing},
  pages={1797--1807},
  year={2018}
}

@article{humaneval,
  title={Evaluating large language models trained on code},
  author={Chen, Mark and Tworek, Jerry and Jun, Heewoo and Yuan, Qiming and Pinto, Henrique Ponde De Oliveira and Kaplan, Jared and Edwards, Harri and Burda, Yuri and Joseph, Nicholas and Brockman, Greg and others},
  journal={arXiv preprint arXiv:2107.03374},
  year={2021}
}

@article{gsm8k,
  title={Training verifiers to solve math word problems},
  author={Cobbe, Karl and Kosaraju, Vineet and Bavarian, Mohammad and Chen, Mark and Jun, Heewoo and Kaiser, Lukasz and Plappert, Matthias and Tworek, Jerry and Hilton, Jacob and Nakano, Reiichiro and others},
  journal={arXiv preprint arXiv:2110.14168},
  year={2021}
}

@inproceedings{mtbench,
  title={Judging llm-as-a-judge with mt-bench and chatbot arena},
  author={Zheng, Lianmin and Chiang, Wei-Lin and Sheng, Ying and Zhuang, Siyuan and Wu, Zhanghao and Zhuang, Yonghao and Lin, Zi and Li, Zhuohan and Li, Dacheng and Xing, Eric and others},
  booktitle={Advances in Neural Information Processing Systems},
  volume={36},
  pages={46595--46623},
  year={2023}
}

@misc{transformersjs,
  author={{Hugging Face}},
  title={{Transformers.js}},
  year={2023},
  url={https://github.com/huggingface/transformers.js}
}

@misc{onnxruntime,
  author={{ONNX Runtime developers}},
  title={{ONNX Runtime}},
  year={2018},
  url={https://onnxruntime.ai}
}

@article{levine2026llamaweb,
  title={Llamas on the Web: Memory-Efficient, Performance-Portable, and Multi-Precision LLM Inference with WebGPU},
  author={Levine, Reese and Sharma, Rithik and Jain, Nikhil and Ramesh, Abhijit and Chen, Zheyuan and Abbas, Neha and Contini, James and Sorensen, Tyler},
  journal={arXiv preprint arXiv:2605.20706},
  year={2026}
}

@inproceedings{wikitext,
  title={Pointer Sentinel Mixture Models},
  author={Merity, Stephen and Xiong, Caiming and Bradbury, James and Socher, Richard},
  booktitle={International Conference on Learning Representations},
  year={2017}
}

@article{smilkov2019tfjs,
  title={Tensorflow. js: Machine learning for the web and beyond},
  author={Smilkov, Daniel and Thorat, Nikhil and Assogba, Yannick and Nicholson, Charles and Kreeger, Nick and Yu, Ping and Cai, Shanqing and Nielsen, Eric and Soegel, David and Bileschi, Stan and others},
  journal={Proceedings of Machine Learning and Systems},
  volume={1},
  pages={309--321},
  year={2019}
}

@inproceedings{ma2019moving,
  title={Moving deep learning into web browser: How far can we go?},
  author={Ma, Yun and Xiang, Dongwei and Zheng, Shuyu and Tian, Deyu and Liu, Xuanzhe},
  booktitle={The World Wide Web Conference},
  pages={1234--1244},
  year={2019}
}

@inproceedings{chen2018tvm,
  title={$\{$TVM$\}$: An automated $\{$End-to-End$\}$ optimizing compiler for deep learning},
  author={Chen, Tianqi and Moreau, Thierry and Jiang, Ziheng and Zheng, Lianmin and Yan, Eddie and Shen, Haichen and Cowan, Meghan and Wang, Leyuan and Hu, Yuwei and Ceze, Luis and others},
  booktitle={USENIX Symposium on Operating Systems Design and Implementation},
  pages={578--594},
  year={2018}
}

@misc{bitgpu,
  author={{stfurkan}},
  title={bitgpu},
  year={2026},
  howpublished={\url{https://github.com/stfurkan/bitgpu}}
}

@article{maczan2026measuring,
  title={Measuring and Reducing WebGPU Dispatch Overhead for LLM Inference},
  author={Maczan, J{\k{e}}drzej},
  journal={arXiv preprint arXiv:2608.08730},
  year={2026}
}

@inproceedings{chen2025weinfer,
  title={WeInfer: Unleashing the power of WebGPU on LLM inference in web browsers},
  author={Chen, Zhiyang and Ma, Yun and Shen, Haiyang and Liu, Mugeng},
  booktitle={Proceedings of the ACM on Web Conference},
  pages={4264--4273},
  year={2025}
}

@inproceedings{jia2024jit,
  title={Empowering in-browser deep learning inference on edge through just-in-time kernel optimization},
  author={Jia, Fucheng and Jiang, Shiqi and Cao, Ting and Cui, Wei and Xia, Tianrui and Cao, Xu and Li, Yuanchun and Wang, Qipeng and Zhang, Deyu and Ren, Ju and others},
  booktitle={Proceedings of the Annual International Conference on Mobile Systems, Applications and Services},
  pages={438--450},
  year={2024}
}

@inproceedings{kwon2023vllm,
  title={Efficient memory management for large language model serving with pagedattention},
  author={Kwon, Woosuk and Li, Zhuohan and Zhuang, Siyuan and Sheng, Ying and Zheng, Lianmin and Yu, Cody Hao and Gonzalez, Joseph and Zhang, Hao and Stoica, Ion},
  booktitle={Proceedings of the Symposium on Operating Systems Principles},
  pages={611--626},
  year={2023}
}

@inproceedings{miao2024specinfer,
  title={Specinfer: Accelerating large language model serving with tree-based speculative inference and verification},
  author={Miao, Xupeng and Oliaro, Gabriele and Zhang, Zhihao and Cheng, Xinhao and Wang, Zeyu and Zhang, Zhengxin and Wong, Rae Ying Yee and Zhu, Alan and Yang, Lijie and Shi, Xiaoxiang and others},
  booktitle={Proceedings of the ACM International Conference on Architectural Support for Programming Languages and Operating Systems, Volume 3},
  pages={932--949},
  year={2024}
}

@inproceedings{li2024eagle,
  title={EAGLE: Speculative Sampling Requires Rethinking Feature Uncertainty},
  author={Li, Yuhui and Wei, Fangyun and Zhang, Chao and Zhang, Hongyang},
  booktitle={International Conference on Machine Learning},
  pages={28935--28948},
  year={2024},
  organization={PMLR}
}

@inproceedings{chen2024sequoia,
  title={Sequoia: Scalable and robust speculative decoding},
  author={Chen, Zhuoming and May, Avner and Svirschevski, Ruslan and Huang, Yuhsun and Ryabinin, Max and Jia, Zhihao and Chen, Beidi},
  booktitle={Advances in Neural Information Processing Systems},
  volume={37},
  pages={129531--129563},
  year={2024}
}

@inproceedings{mamou2024dynamic,
  title={Dynamic Speculation Lookahead Accelerates Speculative Decoding of Large Language Models},
  author={Mamou, Jonathan and Pereg, Oren and Korat, Daniel and Berchansky, Moshe and Timor, Nadav and Wasserblat, Moshe and Schwartz, Roy},
  booktitle={Conference on Neural Information Processing Systems Efficient Natural Language and Speech Processing Workshop},
  pages={456--467},
  year={2024},
  organization={PMLR}
}

@inproceedings{huang2025specdecpp,
  title={SpecDec++: Boosting Speculative Decoding via Adaptive Candidate Lengths},
  author={Huang, Kaixuan and Guo, Xudong and Wang, Mengdi},
  booktitle={Conference on Language Modeling},
  year={2025}
}

@inproceedings{li-etal-2024-eagle,
  title={Eagle-2: Faster inference of language models with dynamic draft trees},
  author={Li, Yuhui and Wei, Fangyun and Zhang, Chao and Zhang, Hongyang},
  booktitle={Proceedings of the Conference on Empirical Methods in Natural Language Processing},
  pages={7421--7432},
  year={2024}
}

@article{liu2024smartspec,
  title={Optimizing speculative decoding for serving large language models using goodput},
  author={Liu, Xiaoxuan and Daniel, Cade and Hu, Langxiang and Kwon, Woosuk and Li, Zhuohan and Mo, Xiangxi and Cheung, Alvin and Deng, Zhijie and Stoica, Ion and Zhang, Hao},
  journal={arXiv preprint arXiv:2406.14066},
  url={https://arxiv.org/abs/2406.14066v1},
  year={2024}
}

@misc{saxena2023pld,
  title={Prompt Lookup Decoding},
  author={Saxena, Apoorv},
  year={2023},
  month={November},
  url={https://github.com/apoorvumang/prompt-lookup-decoding/}
}

@article{yang2023llma,
  title={Inference with reference: Lossless acceleration of large language models},
  author={Yang, Nan and Ge, Tao and Wang, Liang and Jiao, Binxing and Jiang, Daxin and Yang, Linjun and Majumder, Rangan and Wei, Furu},
  journal={arXiv preprint arXiv:2304.04487},
  year={2023}
}

@inproceedings{he2024rest,
  title={Rest: Retrieval-based speculative decoding},
  author={He, Zhenyu and Zhong, Zexuan and Cai, Tianle and Lee, Jason and He, Di},
  booktitle={Proceedings of the Conference of the North American Chapter of the Association for Computational Linguistics: Human Language Technologies},
  pages={1582--1595},
  year={2024}
}

@inproceedings{fu2024lookahead,
  title={Break the Sequential Dependency of LLM Inference Using Lookahead Decoding},
  author={Fu, Yichao and Bailis, Peter and Stoica, Ion and Zhang, Hao},
  booktitle={International Conference on Machine Learning},
  year={2024}
}

@inproceedings{luo-etal-2025-turning,
  title={Turning trash into treasure: Accelerating inference of large language models with token recycling},
  author={Luo, Xianzhen and Wang, Yixuan and Zhu, Qingfu and Zhang, Zhiming and Zhang, Xuanyu and Yang, Qing and Xu, Dongliang},
  booktitle={Proceedings of the Annual Meeting of the Association for Computational Linguistics},
  pages={6816--6831},
  year={2025}
}

@inproceedings{hu-etal-2025-sam,
  title={Sam decoding: Speculative decoding via suffix automaton},
  author={Hu, Yuxuan and Wang, Ke and Zhang, Xiaokang and Zhang, Fanjin and Li, Cuiping and Chen, Hong and Zhang, Jing},
  booktitle={Proceedings of the Annual Meeting of the Association for Computational Linguistics},
  pages={12187--12204},
  year={2025}
}

@inproceedings{oliaro2024suffix,
  title={Suffixdecoding: Extreme speculative decoding for emerging ai applications},
  author={Oliaro, Gabriele and Jia, Zhihao and Campos, Daniel and Qiao, Aurick},
  booktitle={Advances in Neural Information Processing Systems},
  volume={38},
  pages={126326--126354},
  year={2025}
}

\clearpage
\appendix

\section{Extended Benchmark}
\label{app:big}

The extended benchmark takes the next 100 prompts of every task (60 for MT-Bench), which are disjoint from the 220 prompts of the main experiments, and repeats the workstation experiments with the same engine. Table~\ref{tab:big} confirms the main results at almost five times the scale. Every speculative output equals the greedy output on all 1,060 prompts of every model, in every configuration. Our runtime is 8.8, 8.6, 5.6, 7.0, 6.1, 5.4 and 4.9$\times$ faster than WebLLM for the seven models, against 8.5, 8.6, 5.5, 7.3, 6.1, 5.2 and 4.8$\times$ on the main benchmark, and session memory gains more than on the main benchmark, between 12\% and 22\%, because the longer sessions give the drafter more history.

\begin{table}[h]
\centering
\caption{Decoding latency (ms) on the extended benchmark of 1,060 prompts per model on the workstation.}
\label{tab:big}
\small
\setlength{\tabcolsep}{3pt}
\begin{tabular}{lrrrrrr}
\toprule
Model & WebLLM & Greedy & GPU Loop & Ours & Memory & Heads \\
\midrule
Qwen2.5-0.5B & 12.61 & 4.61 & 2.10 & 1.43 & 1.26 & 1.33 \\
Qwen2.5-1.5B & 19.25 & 6.06 & 3.43 & 2.24 & 1.91 & 1.88 \\
Llama-3.2-1B & 8.62 & 4.02 & 1.90 & 1.55 & 1.30 & 1.42 \\
Qwen2.5-3B & 25.67 & 7.26 & 4.98 & 3.65 & 2.83 & 2.69 \\
Llama-3.2-3B & 18.41 & 6.19 & 3.76 & 3.03 & 2.60 & 2.37 \\
Qwen2.5-7B & 25.19 & 9.29 & 6.12 & 4.67 & 3.93 & 3.57 \\
Llama-3.1-8B & 24.34 & 9.44 & 6.10 & 4.92 & 3.94 & 3.75 \\
\bottomrule
\end{tabular}
\end{table}

\section{Verification Cost Curves}
\label{app:tn}

\begin{figure}[h]
\centering
\includegraphics[width=\columnwidth]{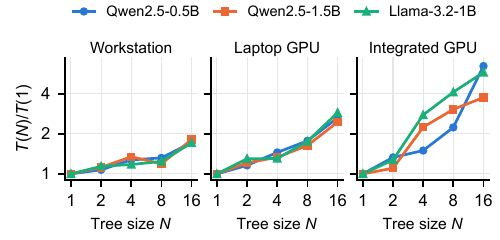}
\caption{Step cost against the tree size, relative to one node.}
\label{fig:tn}
\end{figure}

Figure~\ref{fig:tn} shows the cost $T(N)$ of one verification step with $N$ tree nodes, relative to a step of one node, as the median of five calibrations of the final engine on each GPU. The curves of the three models nearly coincide on the workstation, where a step of 16 nodes costs less than twice a step of one node, so wide trees are cheap. On the laptop GPU, the cost grows steadily, and on the integrated GPU, it rises steeply beyond two nodes, to 3.8 to 6.5 times the cost of one node at 16 nodes. This is the cost structure that the controller measures at load time and that makes a fixed tree of 8 nodes a poor choice on the integrated GPU.

\section{Acceptance Simulation}
\label{app:accept}

\begin{figure}[h]
\centering
\includegraphics[width=\columnwidth]{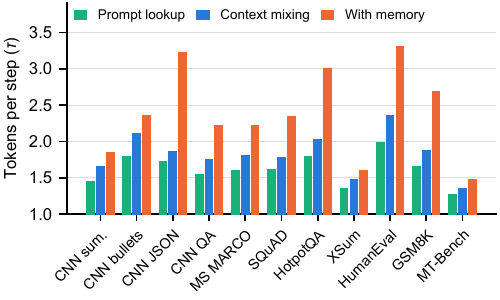}
\caption{Tokens per verification step for each task.}
\label{fig:tasks}
\end{figure}

The speed of speculation rests on how many drafted tokens the model accepts, so we measure acceptance over the full workload in a simulation that replays the greedy outputs of all nine models on all 11 tasks. A step accepts the longest drafted path that matches the greedy continuation, which is exactly what lossless verification accepts in the browser. Prompt lookup decoding with 8 nodes produces 1.62 tokens per step on average, a suffix tree with 16 nodes produces 1.73, and our context-mixing drafter produces 1.78 with 16 nodes and 1.83 with 32 nodes. Figure~\ref{fig:tasks} shows that context mixing leads on every task. Its advantage is largest where the output reuses spans of the input under varying orders, such as the bullet points over news articles and the code completions of HumanEval, because blending the continuations of all match orders lets a short match still propose a confident token.

Session memory changes the picture more than the choice of drafter. Context mixing with 32 nodes rises from 1.83 to 2.40 tokens per step when the drafter can use the outputs of earlier requests in the same session, and the gain holds for every model, from 2.10 for Qwen2.5-1.5B to 2.60 for Phi-3.5-mini. The gain is largest for structured outputs, where tokens per step rise from 1.86 to 3.23 for JSON extraction, from 2.37 to 3.32 for HumanEval and from 1.89 to 2.69 for GSM8K, because consecutive answers share their format, their phrasing and their reasoning templates. It is smallest for open-ended chat and single-sentence summaries, whose outputs share little with each other. The memory costs nothing at prefill, since it enters only the scan of the drafter and never the KV cache.

\begin{figure}[H]
\centering
\includegraphics[width=\columnwidth]{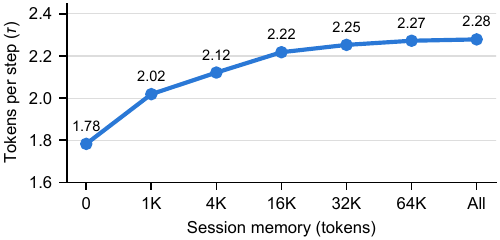}
\caption{Tokens per step against the size of the session memory, with a budget of 16 nodes.}
\label{fig:memory}
\end{figure}

\section{Session Memory Size}
\label{app:memory}

Figure~\ref{fig:memory} replays the outputs of all nine models with session memories of growing size, where All keeps every earlier output of the session. The number of tokens per step rises quickly up to a few thousand tokens of memory and then flattens. The 32K tokens that the runtime keeps in a GPU buffer reach 2.25 tokens per step, against 2.27 with 64K tokens and 2.28 with every earlier output, so this budget captures 95\% of the gain of an unlimited memory.

\end{document}